\documentclass[journal=apchd5,manuscript=article]{achemso}

\SectionNumbersOn
\usepackage{graphicx}% Include figure files
\usepackage{dcolumn}% Align table columns on decimal point
\usepackage{bm}% bold math
\usepackage{color}
\usepackage{soul}
\usepackage{graphicx}
\usepackage{verbatim}
\usepackage{latexsym}
\usepackage{setspace}
\usepackage[utf8]{inputenc}
\usepackage[english]{babel}
\usepackage{amsmath}
\usepackage{amssymb}
\usepackage{xcolor}
\usepackage{mathtools}

\newcommand{\nc}{\newcommand}
\nc{\nn}{\nonumber}

\def\e{\mathcal{E}}

\newcommand{\vast}{\bBigg@{3}}
\newcommand{\Vast}{\bBigg@{5}}
\newcommand{\XYZ}{\bf }

\usepackage[version=3]{mhchem} % Formula subscripts using \ce{}

\author{Ieng-Wai Un}
\affiliation[SECE-BGU]
{School of Electrical and Computer Engineering, Ben-Gurion University of the Negev, Ben-Gurion University}
\email{iengwai@bgu.ac.il}

\author{Yonatan Sivan}
\affiliation[SECE-BGU]
{School of Electrical and Computer Engineering, Ben-Gurion University of the Negev, Ben-Gurion University}
\title{The role of heat generation and {\XYZ fluid} flow in plasmon-enhanced reduction-oxidation reactions}

\abbreviations{IR,NMR,UV}

\begin{document}

%\begin{tocentry}

%\end{tocentry}

\begin{abstract}
Recently, we have shown that thermal effects play a crucial role in speeding up the rate of bond-dissociation reactions. This was done by applying a simple temperature-shifted Arrhenius Law to the experimental data, corroborated with detailed account of the heat diffusion occurring within the relevant samples and identification of errors in the temperature measurements. Here, we provide three important extensions of our previous studies. First, we analyze thermal effects in reduction-oxidation (redox) reactions, where charge transfer is an integral part of the reaction. Second, we analyze not only the spatial distribution of the temperature, but also its temporal dynamics. Third, we also model the fluid convection and stirring. An analysis of two exemplary experimental studies allows us to show that thermal effects can explain the experimental data in one of experiment (Baumberg and coworkers), but not in the other (Jain and coworkers), showing that redox reactions are not necessarily driven by non-thermal charge carriers.
\end{abstract}

\section{Introduction}

The idea of using illuminated metallic surfaces to enhance the yield of chemical reactions (aka plasmon-assisted photocatalysis) was first proposed by Nitzan and Brus in 1981~\cite{Nitzan_PAPC_1981} and experimentally implemented two years later by Chen and Osgood~\cite{Osgood_PAPC_1983}.
Then, after several decades of additional demonstrations (see e.g., Refs~\citenum{Mirkin_PAPC_2001,Brus_PAPC_2003,Tatsuma_2005,plasmonic_photo_synthesis_Misawa_2008,Kitamura_PAPC_2009,Quidant_PAPC_2012}), this line of research has rapidly gained popularity due to several high impact publications (see, e.g. Refs~\citenum{plasmonic_photocatalysis_Clavero,plasmonic-chemistry-Baffou,Valentine_hot_e_review} for some recent reviews). The main motivation behind this growing interest was the claim that the reaction rate can be increased due to the excitation of high energy non-thermal electrons in the metal (aka ``hot'' electrons), which then tunnel out of the metal, and provide the necessary energy for the reactants to allow them to be converted into the products more efficiently. 

However, the claims in some of the more famous papers on the topic~\cite{plasmonic_photocatalysis_1,Halas_dissociation_H2_TiO2,plasmonic_photocatalysis_Linic,Halas_H2_dissociation_SiO2,Halas_Science_2018} were shown to be based on a series of technical errors (including improper temperature measurements, improper data normalization etc., see discussion in Refs.~\citenum{anti-Halas-Science-paper,R2R,Y2-eppur-si-riscalda,Baffou-Quidant-Baldi}), and instead, a purely thermal mechanism was shown to be able to explain the experimental data quite convincingly~\cite{anti-Halas-Science-paper,Y2-eppur-si-riscalda,Dubi-Sivan-APL-Perspective}. In particular, a shifted Arrhenius Law for the reaction rate, $R \sim \exp\left(- \e_a / k_B (T({\bf r}) + a I_{\textrm{inc}}\right)$ whereby the temperature of the system was corrected for the illumination-induced heating was shown to provide an excellent fit to the published data, essentially with no fit parameters. This result was corroborated with the first ever complete calculation of the steady-state electron non-equilibrium in the metal~\cite{Dubi-Sivan}, a consequent Fermi golden-rule argument~\cite{Dubi-Sivan-Faraday} that implied on the unlikeliness of nonthermal electrons to cause the catalysis, and by detailed thermal simulations where the dynamics of the heat generated from each of the nanoparticles (NPs) in the system was properly modelled~\cite{Y2-eppur-si-riscalda}. 

In a consequent paper~\cite{Un-Sivan-sensitivity}, it was shown that the tedious modelling of the contributions of each of the heated NPs in the system can be accurately replaced by an effective medium approximation whereby the penetration depth of the electromagnetic fields was calculated from the NP density and absorption cross-section, and consequently used to describe the heat source. This approach also enabled accounting for the exact reactor geometry, thus, facilitating a quantitative comparison to experimental data. This series of work thus established thermal simulations as a necessary tool in investigations of the underlying physics of plasmon-assisted photocatalysis experiments. % $\delta_{skin}$ - always calculated using the EMT for the EM waves (dilute composite)

The bottom line of the thermal modelling was that when attempting to quantitatively separate thermal and non-thermal effects in plasmom-assisted photocatalysis experiments, one has to overcome a conceptual difficulty - the thermocatalysis control experiments must reproduce the exact spatially non-uniform temperature profile induced by the illumination, otherwise, when subtracting the thermocatalysis rate from the photocatalysis rate (e.g., as in Refs.~\citenum{Halas_Science_2018,Liu-Everitt-Nano-research-2019}), any difference between the temperature distributions in an inaccurate control and the corresponding photocatalysis experiment is bound to be incorrectly interpreted as ``hot'' electron action. This issue is particularly important because the Arrhenius Law shows that the reaction rate has an exponential sensitivity to the temperature distribution. 

Detailed measurements and/or calculations of the temperature distribution in the studied samples indeed constituted a central role in several recent demonstrations of non-thermal effects in plasmon-assisted photocatalysis~\cite{Baldi-ACS-Nano-2018,Liu-Everitt-Nano-research-2019,Cortes_Nano_lett_2019,Boltasseva_LPR_2020}. However, while the simple thermal calculations as done so far were sufficient for relatively simple scenarios, they may not be sufficient to account for more complicated ones. Those include, in particular, high intensity illumination which invokes nonlinear thermo-optic effects~\cite{Donner_thermal_lensing,Sivan-Chu-high-T-nl-plasmonics,Gurwich-Sivan-CW-nlty-metal_NP,IWU-Sivan-CW-nlty-metal_NP}, charge transport (in particular, in electrochemistry experiments~\cite{Tatsuma_2004,Willets_electrochemistry_2018,Cortes_Nano_lett_2019,Caleb_Hill_Heating_Electrochemical_JPCC_2019,Nguyen-Merced-redox,Lange_2020}), or in the presence of strong gas flows, liquid stirring etc.. In this context, % it became important to understand the role of heat convection/diffusion/gas flow / liquid stirring - 
there are several claims in the literature that the latter can effectively eliminate thermal gradients, and thus, make thermocatalysis control experiments based on external heating far more similar to the generally more complex temperature distribution in photocatalysis experiments. As noted above, this can simplify the distinction between thermal and non-thermal effects, however, somewhat peculiarly, to the best of our knowledge, no estimates nor proper numerical simulations of the effectiveness of gas flow, convection or liquid stirring were previously presented. In Ref.~\citenum{Y2-eppur-si-riscalda} we offered some simplistic estimates which showed that gas flow is not expected to be significant in removing temperature gradients; however, one may ask whether the situation changes in a liquid environment, especially in the presence of rapid external stirring which can be stronger than gas flow. 

In order to quantify these effects, in the current manuscript we add to the (validated Ref.~\citenum{Un-Sivan-sensitivity}) effective medium modelling used previously some more advanced effects such as gas flow and liquid convection/stirring via the proper fluid mechanics equations. We reproduce numerically the temperature distribution in two experiments performed in a liquid environment where natural convection and forced stirring occurred. This specific choice of experiments allowed us also to address another common claim, namely, the validity of the thermal analysis to a wider variety of chemical reactions. Indeed, all the chemical reactions we previously identified as likely to be driven by thermal effects in Refs.~\citenum{Y2-eppur-si-riscalda,anti-Halas-Science-paper} were bond-dissociation type. However, a different type of reactions, namely, reduction-oxidation (redox) reactions are, by definition, associated with charge transfer, so that they are frequently associated with the optical generation of ``hot'' electrons or electron-hole pairs~\cite{Jain_viewpoint}. 

The manuscript is organized as follows. In Section~\ref{sec:model} we introduce our improved thermal model along with the fluid dynamics equations. In Section~\ref{sec:experiments} we describe the two experiments, the corresponding temperature gradients and temporal dynamics, and draw conclusions regarding the effectiveness of the convection/stirring as well as the mechanism responsible for the faster reactions reported. Potentially counter-intuitively, we show that % in both cases the convection/stirring does not eliminate the thermal gradients, such that significant differences in the temperature distribution in the photocatalysis and control thermocatalysis remain. In addition, we find that 
the results of the first experiment can be explained just with a thermal theory (i.e., the shifted-Arrhenius model), while the results of the latter cannot.

\section{Model} \label{sec:model}
Previously in Refs.~\citenum{anti-Halas-Science-paper,Y2-eppur-si-riscalda}, calculations of the temperature distribution in plasmon-assisted photocatalysis systems relied on a straightforward solution of the heat equation accounting for the multiple absorbing metal NPs as nanometric heat sources within an effective medium model for the light absorption. Specifically, one solved
\begin{equation}
c_p \rho_0 \dfrac{\partial T({\bf r},t)}{\partial t} = \nabla \cdot \left(\kappa \nabla T\right) + p_\textrm{abs}({\bf r},t), \label{eq:heat-transfer-old}
\end{equation}
where $c_p$ is the heat capacity at constant pressure, $\rho_0$ is the (constant) mass density of the fluid, $T$ is the temperature and the term $\nabla \cdot (\kappa\nabla T)$ accounts for heat diffusion/conduction. Typically, homogenized values of $c_p$, $\rho_0$ and $\kappa$ for the host~\cite{Y2-eppur-si-riscalda} were used; this means that the metal parameters were ignored due to the small volume fraction they occupy. Finally, $p_\textrm{abs}$ is the heat source density induced by the light absorption; it has been shown that by applying the effective medium approximation for the electromagnetic properties, it is well described by~\cite{Un-Sivan-sensitivity} % we ignore here the local field enhancement because this calculation assumes effective medium properties
\begin{align}\label{eq:heat_source}
p_\textrm{abs}({\bf r},t) = \left(I_\textrm{inc}(\varrho)/\delta_\textrm{skin}\right)\textrm{exp}(-|z|/\delta_\textrm{skin}),
\end{align}
where $\varrho$ is the distance from the propagation optical axis, $|z|$ is the distance along the propagation direction of the incident beam from the spot area, $I_\textrm{inc}(\varrho)$ is the intensity profile of the incident beam %$R_b$ is the beam spot radius %{\bf IW - this should be marked somehow in fig. 1} 
and $1/\delta_\textrm{skin}$ is the absorption coefficient experienced by the incident beam; it can either be obtained by the measurement directly (see e.g., Section~\ref{sec:Baumberg-Faraday}), or can be related to the NP density $n_\textrm{p}$ (number per volume) and the absorption cross-section $\sigma_\textrm{abs}$ by using the effective medium theory~\cite{Y2-eppur-si-riscalda,Un-Sivan-sensitivity} (see e.g., Section~\ref{sec:Jain-natcommun})
\begin{align}\label{eq:skin-depth}
1/\delta_\textrm{skin} = n_p\sigma_\textrm{abs}.
\end{align}

In the current study, in order to model more complicated scenarios in which the temperature dynamics is accompanied by significant fluid density and velocity dynamics, we add to the heat equation a term responsible for heat convection and determine the mass velocity and density via the Navier-Stokes equation~(\ref{eq:Navier-Stokes}) coupled with the continuity equation~(\ref{eq:continuity}), namely,  %under the Boussinesq approximation. %{\bf this is an approximation of 1 and 2 only, right?}. 
%In this approximation, it is assumed that the density variation has negligible effects on the flow field, except that it causes the buoyancy force, namely, %{\bf IW - if $p$ is a variable, then, should we still use $c_p$? the choice of parameters ignores the presence of the metal NPs? or uses an EMT calculation as in~\cite{Y2-eppur-si-riscalda}?}
\begin{align}
c_p \rho({\bf r},t) \left[\dfrac{\partial T({\bf r},t)}{\partial t} + ({\bf u}({\bf r},t) \cdot \nabla)T\right] &= \nabla \cdot \left(\kappa \nabla T\right) + p_\textrm{abs}({\bf r},t).\label{eq:heat-transfer} \\
\rho \dfrac{\partial {\bf u}}{\partial t} + \rho({\bf u}\cdot \nabla){\bf u} &= - \nabla p + \rho_0 {\bf g} + (\rho - \rho_0){\bf g} + \nabla \left[\mu\left(\nabla{\bf u} + (\nabla{\bf u})^\intercal\right)\right], \label{eq:Navier-Stokes}\\
\dfrac{\partial \rho}{\partial t} + \nabla\cdot \left(\rho {\bf u}\right) &= 0. \label{eq:continuity}
\end{align}
% {\bf did we neglect $\dot{\rho}?$} 
As well known, these equations ensure conservation of energy, momentum and mass in the fluid. Here, the unknown variables ${\bf u}$, $p$, $\rho$ and $T$ are, respectively, the flow velocity, the pressure, fluid density and temperature. As before, $\rho_0$ is the (constant and uniform) fluid density at the ambient conditions. Since Eq.~(\ref{eq:heat-transfer})-(\ref{eq:continuity}) contain five equations in terms of six unknown flow-field variables, we also require the equation-of-state to relate the fluid density to the pressure and temperature. For liquids, the fluid density can be assumed to be pressure-independent due to the small compressibility of most liquids. This provides the necessary sixth equation. 

%Here, the unknown variables ${\bf u}$, $p$, $\rho$ (and $T$) are, respectively, the flow velocity, the pressure, fluid density (and the temperature). As before, $\rho_0$ is the (constant and uniform) fluid density at the ambient conditions. Since typical experiments of plasmon-assisted photocatalysis were performed at atmospheric pressure, the additional pressure at a depth of a few cm of water (a few hundreds Pa) has negligible effect on the material parameters; $\rho$, $\mu$, $c_p$ and $\kappa$ are, thus, assumed to be pressure-independent. Similarly, we limited ourselves to modest heating levels, in which case it is justified to ignore any temperature dependence of the parameters; this assumption can be relaxed, as e.g., in~\cite{Sivan-Chu-high-T-nl-plasmonics,IWU-Sivan-CW-nlty-metal_NP}.

On the right hand side of Eq.~(\ref{eq:Navier-Stokes}), the terms $-\nabla p$ and $\rho{\bf g}$ are, respectively, the pressure forces and the gravity forces. These two terms nearly cancel each other for liquids. However, the variations of density with temperature which arise in the gravity force give rise to a buoyancy force lifting the fluid. This force is represented by $(\rho-\rho_0){\bf g}$ and is responsible for the natural convection in the fluid. The last term $\nabla \left[\mu\left(\nabla{\bf u} + (\nabla{\bf u})^\intercal\right)\right]$ represents the viscous forces ($\mu$ is the viscosity). This is usually small when only the natural convection in the fluid is considered.  

In addition, for plasmon-assisted photocatalysis in liquid phase, since typical experiments were performed at atmospheric pressure, the additional pressure at a depth of a few cm of liquid (a few hundreds Pa) has a negligible effect on the material parameters; $\mu$, $c_p$ and $\kappa$ are, thus, assumed to be pressure-independent. However, the temperature dependence of these parameters is non-negligible.

The above assumptions are suitable for liquid hosts. One can also use the model equations~(\ref{eq:heat-transfer})-(\ref{eq:continuity}) to calculate the temperature distribution in samples in which the catalysts are nanostructured solids and the reactants are gases (e.g., as in Refs.~\citenum{Halas_dissociation_H2_TiO2,Halas_H2_dissociation_SiO2,Halas_Science_2018,plasmonic_photocatalysis_Linic}). In these photocatalysis experiments, the catalyst samples are usually put in a reaction chamber consisting of inlet and outlet channels between which gases flow in the chamber. When modeling these systems, we have to take into account the pressure-dependence of the gas density via the equation of state due to the high compressibility of gas. To include the gas flow, one should use the inlet boundary condition and specify the gas velocity in the inlet as $\left(=\dfrac{\textrm{volumetric flow rate}}{\textrm{inlet section area}}\right)$ via a Dirichlet boundary condition. Simple estimates of the power flow due to convection performed in Ref.~\citenum{Y2-eppur-si-riscalda} showed that such gas flows can have, at most, a moderate quantitative effect on the temperature distribution in the system. More accurate modelling requires knowledge (or computation) of the gas flow velocity inside the sample; this information is not usually available, and is hard to compute.

\section{Representative Analysis}\label{sec:experiments}
In this Section, we apply the model described above to two representative experiments. In contrast to a variety of experimental studies that aim to mimic practical conditions for catalysis, these two studies chose a simple configuration that facilitates the isolation of the underlying physical mechanism. In particular, in both experiments the heat source was highly localized, such that the spatio-temporal dynamics of heat occurred in a homogeneous environment, and thermal gradients and dynamics were pronounced.

\subsection{Huang {\em et al.} [J. Baumberg's group,~\citenum{Baumberg-Faraday}]}\label{sec:Baumberg-Faraday}
\begin{figure}[t]
\centering{
\includegraphics[width=1\textwidth]{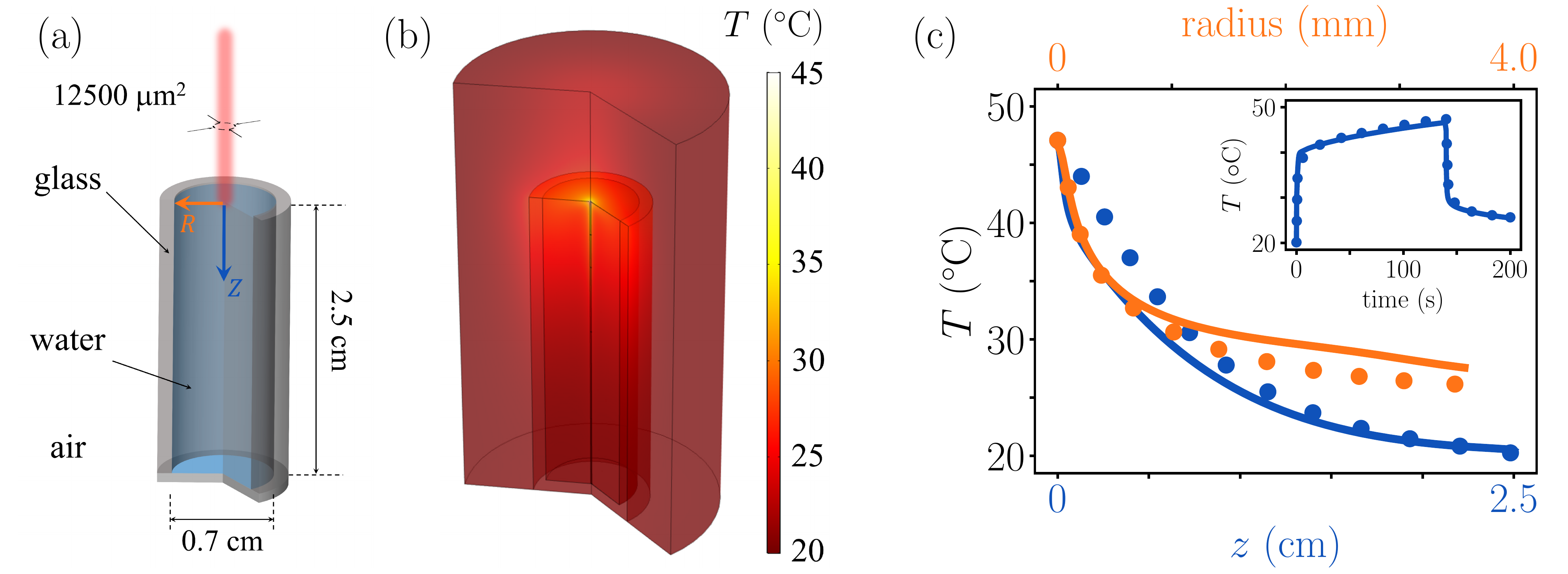}
\caption{(Color online) (a) Schematic of the simulation domain. A laser beam of wavelength 785 nm is focused at the water–air interface from the top of the open cylindrical glass cuvette. (b) The temperature distribution at $t = 140$ s. (c) The temperature distribution along the cylindrical axis (blue solid line) and along the radius at the water–air interface (orange solid line) at $t = 140$ s. The inset shows the temperature dynamics at the laser spot center. The dotted symbols represent the corresponding temperature distribution (and temperature dynamics in the inset) when the free convection of water is neglected.} \label{fig:Baumberg_plot}}
\end{figure}

First, we look at the results of Huang {\em et al.}~\cite{Baumberg-Faraday} who studied dithionite-mediated redox reactions in a water suspension of Au NP aggregates. The aggregates were estimated to consist of a random arrangement of 15 NPs on average, each $\sim 60$nm in size. The suspension was put in a 1 cm $\times$ 0.4 cm $\times$ 2.5 cm cuvette which was heated using a $785$ nm spatially-narrow ($\sim 110\ \mu$m diameter) 125 mW CW beam. The presence of the Au NPs enhances not only the reaction, but also the Raman signal from the redox reporter molecules, allowing for surface enhanced Raman spectroscopy (SERS) to be used to monitor the rate of the reaction (i.e., the density of both reactants and products) taking place in the illuminated region, see Fig.~\ref{fig:Baumberg_plot}(a). No reaction takes place in the absence of the metal NPs.

Initially, the cuvette was subject to a hot water bath whose temperature was externally controlled. It was then observed that products were detected for heating of the sample by 8 K or more above room temperature. This enabled Huang {\em et al.} to conclude that if under illumination the sample temperature remained lower than that threshold, but reactants can be detected, then, the reaction is likely to be driven by high energy non-thermal electrons.

A rough estimate of the heat generation and transfer to the suspension from each {\em single} aggregate then allowed Huang {\em et al.} to predict that the (steady-state) temperature rise of a single aggregate with respect to its environment may reach $1-2$ K. In addition, they performed an estimate of the {\em overall} cuvette temperature rise which relied on the assumption of infinitely fast heat diffusion (such that the total heat generated in the sample spreads evenly in its volume). This gave a slightly higher temperature of $\sim 4$ K for the average temperature in the sample (due to the accumulated heating from the multiple aggregates). With these conservative estimates and in light of the difficulties to measure the temperature distribution at high accuracy, the authors were unable to ascribe clearly the many effects observed to non-thermal carriers.

Here, we employ the model~(\ref{eq:heat-transfer})-(\ref{eq:continuity}) to improve upon the original heating estimates. Our motivation to do so is the fact that since the heat diffusion has a finite strength, it is reasonable to assume that some level of non-uniformity exists in the suspension, such that the temperature of the illuminated (and probed) region may be higher than its surroundings. Thus, in order to determine the actual level of temperature non-uniformity, we first perform a standard heat transfer simulation (neglecting the natural convection in water, namely, we solve Eq.~(\ref{eq:heat-transfer}) by assuming that ${\bf u} = 0$, but accounting for heat diffusion in water) with a heat source Eq.~\eqref{eq:heat_source} in which the transverse intensity profile has a Gaussian shape, namely, $I_\textrm{inc}(\varrho) = \left(P_\textrm{inc}/(\pi \varrho_b^2)\right) \exp\left(-\varrho^2/\varrho_b^2\right)$, where $P_\textrm{inc}$ is the illumination power and $\varrho_b$ is the beam spot radius. We note, however, that the 3D simulation of fluid flow and the temperature dynamics of the water in a rectangular cuvette requires a large computing time. To obtain a quick result without compromising accuracy, we replaced the rectangular cuvette used in the experiment by a cylindrical one, such that the height and the cross-section of the model are, respectively, equal to the height and the cross-section of the actual cuvette used in the experiments, see Figure~\ref{fig:Baumberg_plot}(a). This allows us to simplify the 3D model into a 2D model using an axisymmetric geometry. As the following result show, the cuvette size and shape have a negligble effect on the dynamics in this case, as the cuvette is sufficiently large.

The analysis below is enabled by the extraction of a key parameter from the experimental data, namely, the absorption coefficient $1/\delta_\textrm{skin}$~(\ref{eq:skin-depth}) which is itself deduced from the attenuation measurement. Specifically, the reported $83\%$ attenuation per $cm$ corresponds to a penetration depth of $\delta_{skin} \approx 5.6$ mm ($\sim 20\%$ of the total thickness). We set the bottom of the cuvette to a temperature of $20^\circ$C; the top surface of the water and the outer surfaces of the cuvette dissipate heat via natural convective flow of air driven by the temperature difference between the cuvette and its surrounding. 

Initially, the system is at the room temperature. Then, the laser is turned on for 140 seconds. The simulation (see Figure~\ref{fig:Baumberg_plot}(b)-(c)) shows a strongly non-uniform temperature profile occurring because the beam is much smaller than the cuvette in which the liquid suspension was. In particular, the temperature in the region close to the illuminated NPs grows by about $27$ K, while regions further away hardly heat up at all.

These hydrostatic results imply that the highly non-uniform temperature distribution induces a density difference between the hot and cold regions thereby leading to buoyancy-driven flow such that the illuminated region cools down. For this reason, we further improve the temperature calculations by modeling both heat diffusion and convection as well as simulating the fluid flow via Eqs.~(\ref{eq:heat-transfer})-(\ref{eq:continuity}). This simulation shows that the temperature non-uniformity induced by the local heating is hardly affected by the natural convection, see Figure~\ref{fig:Baumberg_plot}(b). In particular, the temperature rise at the illumination region decreases by only a few K due to the natural convection, see Figure~\ref{fig:Baumberg_plot}(c). The lower part of the sample is even less affected, because its bottom is set to room temperature. 

Thus, since the temperature of the illuminated (hence, monitored) region is much higher than the $8$ K threshold, we conclude that the reaction in Ref.~\citenum{Baumberg-Faraday} was likely driven by the heat generated in the sample (rather than by generation of high energy non-thermal carriers).

This conclusion is further supported by additional evidence that the temperature {\em dynamics} that emerges from our simulations is consistent with the experimentally-measured {\em product} dynamics. Such an analysis was attempted in Ref.~\citenum{Frontiera-2018}, yet, was applied to heating of a single NP by a single pulse; thus, much like Ref.~\citenum{plasmonic_photocatalysis_Linic}, it overlooked cumulative thermal effects and reached an invalid conclusion about the negligible role played by thermal effects in Ref.~\citenum{Frontiera-2018}. In contrast, the simulation results (see inset of Figure~\ref{fig:Baumberg_plot}(c)) which look at the macroscopic (i.e., cumulative) heat dynamics show that the temperature of the illuminated region increases rapidly by $\sim 20$ K within the first second of the illumination (before settling on a much slower heating rate). To see that, it is insightful first to understand the origin of the time scale for the temperature rise in our simulations. This is done by approximating the heat source in the experiment as an infinitely small heat source turned on abruptly, namely, $P_\textrm{inc}\delta^{(3)}({\bf r})H(t)$. The ensuing heat dynamics is given by 
\begin{align}\label{eq:temp_infty_sm}
\Delta T({\bf r},t) \approx \Delta T({\bf r},\infty) \left[1 - \textrm{erf}\left(\dfrac{r}{\sqrt{4Dt}}\right)\right],
\end{align} 
where $\Delta T({\bf r},\infty)$ is the ``steady-state'' temperature rise (i.e., at long times, assuming that the laser was not turned off), $D = \kappa / c_p\rho_0$ is the (effective) thermal diffusivity, $\textrm{erf}(x)$ is the error function and $H(t)$ is the Heaviside step function. 

As shown in the inset of Fig.~\ref{fig:Baumberg_plot}(c), within roughly 1 second, the temperature rise reaches about $80\%$ of its ``steady-state'' value before slowing down significantly. 

This is the time scale necessary for the build-up of the temperature gradient for which the outgoing heat flux becomes comparable to the illumination-induced heat generation. In that sense, this dynamics is independent of the geometry of the cuvette, i.e., it is determined only by the beam size and the host diffusivity. Then, based on the estimate $\textrm{erf}(x) \sim x$, we need to find the time that satisfies $\dfrac{r}{\sqrt{4 D t}} \sim 0.2$. Using the water diffusivity ($\sim 0.15 \cdot 10^{-6} m^2/s$) and the beam radius, we get that the temperature rises within the time scale of $\sim 0.5$ s. 

This temperature dynamics matches with the experimental observation that the characteristic bands of the products appeared in the SERS emission after a few seconds of illumination. Due to the instantaneous nature of the SERS measurement technique, the limitations discussed in Ref.~\citenum{Baffou-Quidant-Baldi} can be alleviated such that the growth of the reaction rate can be directly ascribed to heat. To our knowledge, this is the first successful attempt to distinguish thermal from nonthermal effects in plasmon-assisted photocatalysis by the time dynamics, as originally suggested by Baffou {\em et al.} in Ref.~\citenum{Baffou-Quidant-Baldi}. Indeed, the (rather unique ) uniformity of the sample and the large and relatively simple reactor geometry (with respect to the illumination beam) in this experiment enables the simple analysis above. Nevertheless, Raman signal dynamics of a few second time scales was observed in various other plasmon-assisted photocatalysis experiments which involved a more complicated reactor geometry, see e.g., Refs.~\citenum{sarhan2019importance,sarhan2019_JPCC}. Although an estimate of the heat dynamics is far more complex in these cases, it is more than reasonable to associate the faster reaction rates to heating, especially since simple estimates we made for these cases imply that the thermal dynamics may be on the few-second time scale or even faster.

A few-second time scale dynamics was also observed in many other experiments where the signal was not monitored instantaneously, but rather, using a mass spectrometer. While this prevents one from drawing firm conclusions about the underlying physical mechanism (see further discussion in Ref.~\citenum{Baffou-Quidant-Baldi}), the slow dynamics in these cases is likely due to thermal effects, especially in those experiments that were shown to be thermally-driven by other means (such as thermal calculations, and fits to Arrhenius curves; e.g., Refs.~\citenum{christopher2011visible,plasmonic_photocatalysis_Linic,Halas_dissociation_H2_TiO2,Liu-Everitt-Nat-Comm-2016,Halas_Science_2018,Halas-Nature-Catalysis-2020,Wuhan_PAPC_2020}). Similarly, few-second time scale dynamics was also observed in various photoelectrochemistry experiments~\cite{Brus_photocatalysis_2007,Willets_electrochemistry_2018,Simpkins_SPR_photocatalysis,Moskovits_Xiamen_2019,Guangzhou_PAPC_2020,Wuhu_PAPC_2020}. While the heat dynamics analysis is applicable to these works too, it is not possible yet to draw firm conclusions before the theory is complemented by a proper study of the charge dynamics.

\subsection{Yu {\em et al.} [P. K. Jain's group,~\citenum{yu2019plasmonic}]}\label{sec:Jain-natcommun}
\begin{figure}[t]
\centering{
\includegraphics[width=1\textwidth]{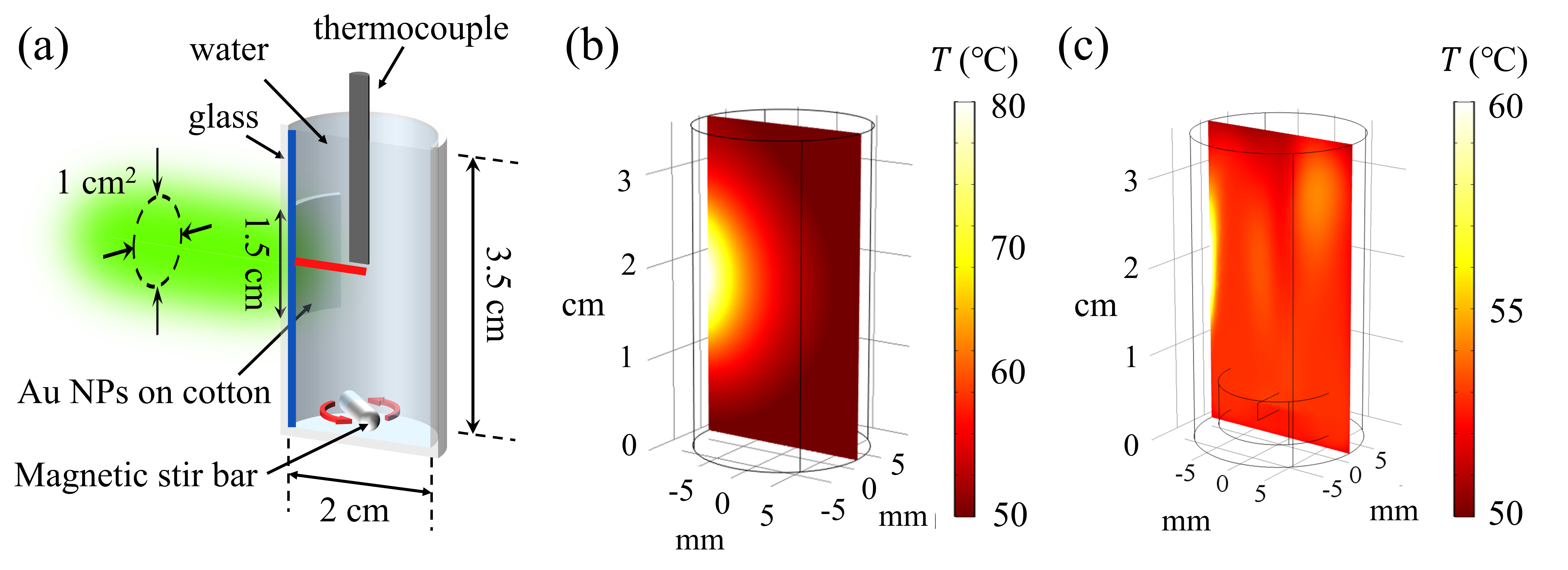}
\caption[]{(Color online) (a) Schematic of the simulation domain. A continuous-wave laser beam is illuminating the side wall of an open cylindrical glass cuvette. A 2D cross-section of the steady-state temperature distribution in the cuvette for (b) $\Omega = 0$ rpm (without stirring) and (c) $\Omega = 120$ rpm. 
} \label{fig:Jain_scheme_map}}
\end{figure}

In contrast, the experiment of Yu {\em et al.}~\cite{yu2019plasmonic} cannot be explained within our simple thermal model shifted-Arrhenius Law, because our thermal calculations predict a much lower reaction rate than what was actually observed experimentally. To see this explicitly, we perform the thermal calculations for the configuration studied in Ref.~\citenum{yu2019plasmonic}. Specifically, the sample in Ref.~\citenum{yu2019plasmonic} consisted of a $3.5$ cm long cuvette with a circular cross-section of radius of $1$ cm, see Fig.~\ref{fig:Jain_scheme_map}(a). The bottom of the cuvette was attached to an external heater, and was complemented with a small rotor that was used to stir the $10$ cc water volume. A rectangular cotton patch was attached to the cuvette wall, and was doped with $\sim 2 \times 10^{13}$ Au NPs of $\sim 12$ nm diameter. The patch was illuminated by a continuous-wave laser beam of spot size $\sim$ 1 cm$^2$ at a wavelength of $532$ nm and incident power of $1$ W. A thermocouple was inserted at the center of the cuvette, half way into its depth to monitor the temperature.

The thermal distribution in this system was chosen as a case study in Ref.~\citenum{Un-Sivan-sensitivity} in order to demonstrate explicitly the dependence of the temperature distribution and total reaction rate on the parameters of the system. These calculations relied on the formalism described in Ref.~\citenum{Y2-eppur-si-riscalda} which consisted of a summation of the heat diffusing from a large ensemble of absorbing metal NPs; the calculation was also validated using an effective medium approximation in Ref.~\citenum{Un-Sivan-sensitivity}.

In the current work, in order to model the experimental conditions more faithfully, we simulate the heat transfer and convection of the liquid heated by the illumination via Eqs.~\eqref{eq:heat-transfer}-\eqref{eq:continuity}. To account for the stirring, we separate the calculation domain into two parts, a rotating part and a stationary part. The rotating part contains the magnetic stir bar itself~\footnote{The specific stir bar geometry we used may not have been identical to the one used in the experiment, yet, this is not expected to have any significant effect on the results. } and the water in its vicinity, while the stationary part contains all other domains outside the rotating part. In the rotating part, we reformulate Eqs.~\eqref{eq:heat-transfer}-\eqref{eq:continuity} in a rotating frame but in terms of the velocity flow ${\bf u}$ and temperature $T$ in the stationary frame, namely,
\begin{align}
c_p \rho \left[\dfrac{\partial }{\partial t} + \left({\bf u} - \bm{\Omega} \times {\bf r}\right) \cdot \nabla\right] T &= \nabla \cdot \left(\kappa \nabla T\right) + p_\textrm{abs}. \label{eq:heat-transfer-rot-rfm} \\
\rho \left[\dfrac{\partial }{\partial t} + \left({\bf u} - \bm{\Omega} \times {\bf r}\right) \cdot \nabla\right]{\bf u} &= - \nabla p + \rho_0 {\bf g} + (\rho - \rho_0){\bf g} + \nabla \left[\mu\left(\nabla{\bf u} + (\nabla{\bf u})^\intercal\right)\right]\label{eq:Navier-Stokes-rot-rfm} \\
\left[\dfrac{\partial }{\partial t} - \left(\bm{\Omega}\times{\bf r}\right)\cdot\nabla\right]\rho + \nabla\cdot \left(\rho {\bf u}\right) &= 0,\label{eq:continuity-rot-rfm}
\end{align}
where $\bm{\Omega} = \Omega {\hat{\bf z}}$ is the rotation vector with $\Omega$ being the angular speed. In the stationary part, the ordinary heat transfer and Navier-Stokes equations~\eqref{eq:heat-transfer}-\eqref{eq:continuity} are solved. For the temperature, we set a continuity condition at the boundary between the rotating and the stationary parts. At the bottom of the cuvette, we set the boundary condition to the controlled temperature, effectively assuming that the cuvette is in perfect contact with the hotplate surface. At the top and outer surfaces, we use a convective heat flux boundary condition driven by the temperature difference between the cuvette and the surrounding. For the flow velocity, we apply the continuity condition to couple the flow velocity on the boundary between the rotating and stationary parts. In the stationary part, no-slip condition is set for the fluid flow at the interface between the liquid and the cuvette while a slip condition is set at the water-air interface.

When the liquid is not stirred, the temperature distribution within the sample is clearly non-uniform, see Fig.~\ref{fig:Jain_scheme_map}(b). In particular, the temperature near the cotton patch is $\sim 30^\circ$C higher than the average temperature in the bulk of the liquid and higher than the measured temperature. 

\begin{figure}[t]
\centering{
\includegraphics[width=0.6\textwidth]{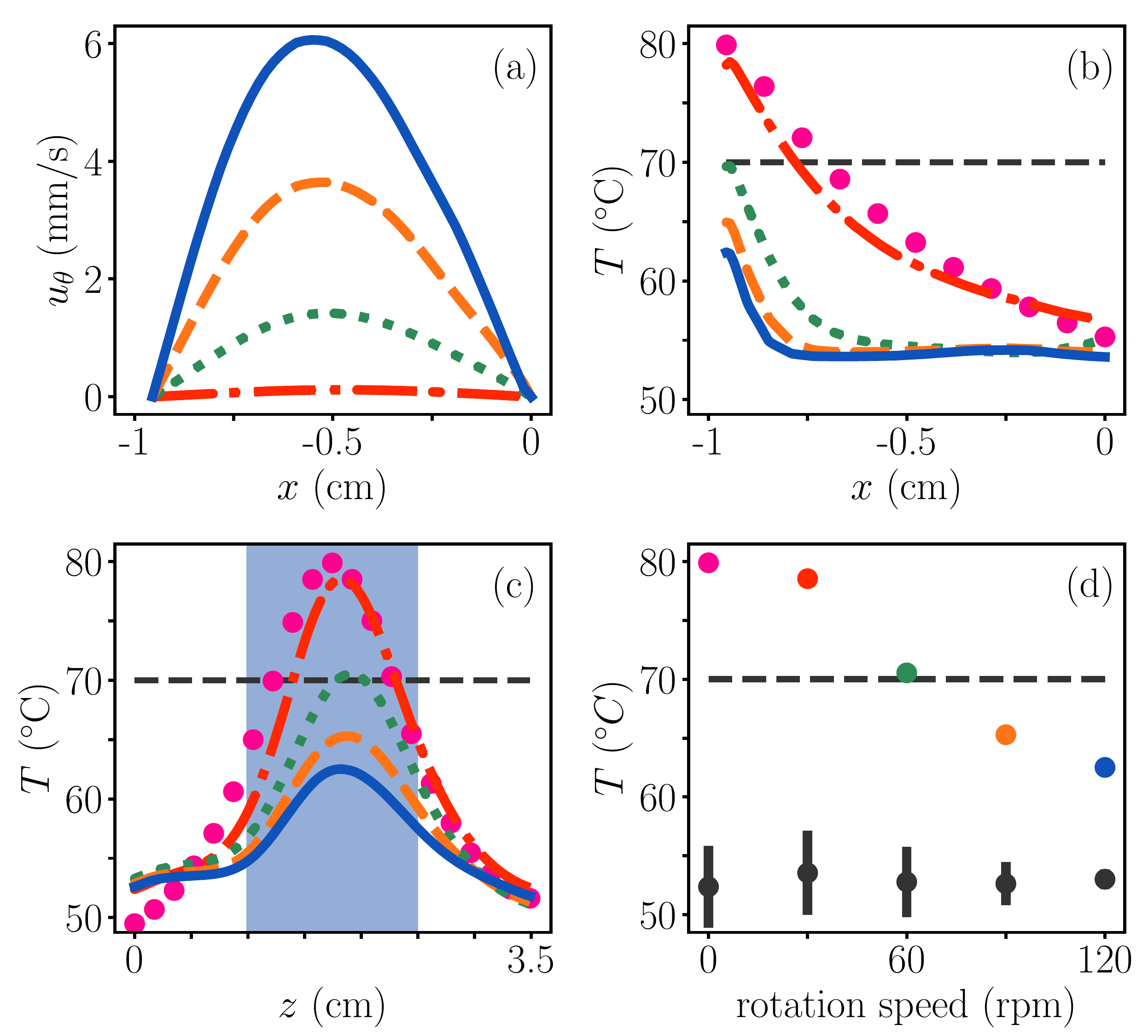}
\caption{(Color online) (a) 1D cross-section (marked by the red solid line in Fig.~\ref{fig:Jain_scheme_map}(a)) of the steady-state azimuthal velocity $u_\theta$ for various angular speeds of the magnetic stir bar (30 rpm, red dotted dash line; 60 rpm, green dotted line; 90 rpm, orange dashed line; 120 rpm, blue solid line). (b) and (c) 1D cross-sections of the steady-state temperature distribution along $x$-direction (at $z = 1.75$ cm; marked by the red solid line in Fig.~\ref{fig:Jain_scheme_map}(a)) and along the $z$-direction (at $x = -1$ cm; marked by the blue solid line in Fig.~\ref{fig:Jain_scheme_map}(a)), respectively, for the same angular speeds of the magnetic stir bar in (a); the case of 0 rpm is added (magenta dots). The shaded region in (c) indicates the NP-coated cotton substrate. (d) The color dots represent the maximum temperature on the cotton patch for various angular speeds of the magnetic stir bar. The black dots represent the averaged temperature at the thermocouple, as appropriate for its high thermal conductivity. The error bars represent the temperature variation along the thermocouple.The black dashed lines in (b), (c) and (d) represent the temperature 70$^\circ$C measured by the thermocouple in the control experiments~\cite{yu-Jain_energy-lett-2019}.}\label{fig:Jain_plot}
}
\end{figure}

The intuitive expectation is that the stirring would reduce these gradients within the sample. Indeed, as shown in Fig.~\ref{fig:Jain_scheme_map}(c) for stirring speed of 120 rpm, the temperature variation in the bulk of the liquid is much smaller. However, since the heat source (namely, the cotton patch) is positioned on the wall of cuvette, where the liquid flow vanishes (due to the no-slip boundary condition), the heat gradients in the vicinity of the heat source are not reduced as efficiently by the stirring as in the bulk of the sample. Indeed, Fig.~\ref{fig:Jain_plot}(a) shows that the azimuthal flow velocity (parallel to the cuvette wall) forms the well-known boundary layer~\cite{Kundu-Fluid-Mech-2016,gerhart2020fundamentals} whereby the flow velocity drops linearly towards the cuvette wall. As a result, heat drainage away from the heat source position is not as effective as in other parts of the sample. In that sense, our results (Fig.~\ref{fig:Jain_plot}(a)) imply that stirring could have been most effective were the heat source positioned half way between the wall and cuvette center, or simply spread evenly within the cuvette volume.

As a result, the radial gradients of the temperature are nearly eliminated at stirring speeds of 120 rpm in most of the volume, yet, they persist near the cuvette wall, see Fig.~\ref{fig:Jain_plot}(b)-(c). More specifically, the maximal temperature (occurring on the cotton patch) is reduced from $\sim 80^\circ$C (no stirring) to $\sim 63^\circ$C (120 rpm). Similarly, the radial temperature gradient is reduced from $\sim 27^\circ$C to $\sim 10^\circ$C in a monotonic yet nonlinear fashion (see also Fig.~\ref{fig:Jain_plot}(d)). Importantly, this shows that the thermocouple position in the experiment causes its reading to underestimate the temperature on the patch where the limiting step of the reaction is expected to occur. 

In the experiment itself, a much higher rotation speed was applied~\cite{Jain-private}. Unfortunately, it is difficult to predict how much would the gradients be further reduced with such an additional increase of the stirring speed. Indeed, 3D maps of the liquid flow (not shown) imply on the development of turbulent flow, an effect which will cause the convective heat transfer to be substantially enhanced by the intense mixing of the fluid~\cite{Cenge-heat-transfer-book}. In this case, a much more sophisticated numerical framework compared to the one employed here is required for the correct treatment of the turbulent convection heat transfer. 

Nevertheless, the authors of Ref.~\citenum{yu2019plasmonic} were careful enough to account for this potential complication. In particular, in order to demonstrate that the product generation in their experiment does not originate from simply a photothermal effect, the authors performed thermal control experiments where the stirred liquid in the cuvette was heated up externally in the dark whereby the thermocouple temperature reached $70^\circ$C~\cite{yu-Jain_energy-lett-2019}. Under this condition, they reported that negligible product generation was observed.

Thus, while it is unclear how much the gradients associated with the boundary layer would be further reduced at higher stirring speeds, already at the rotation speeds we could reliably simulate it seems that the temperature of the NPs is lower than in the $70^\circ$ control experiment. Thus, since the thermocatalysis experiment yielded no reaction, thermal effects seem to be incapable of explaining the experimental observation of the reactants under illumination, and in fact, optically-induced non-thermal electrons seem to be responsible for the reaction rate in its entirety; a similar situation was demonstrated in Ref.~\citenum{Boltasseva_LPR_2020}. In fact, together with the quantitative predictions of the number of nonthermal electrons generated in the metal~\cite{Dubi-Sivan,Dubi-Sivan-Faraday}, with a quantification of the number of nonthermal electrons that participated in their reaction and with reactant diffusion dynamics (needed to estimate the arrival time of the reactants to the gas chromatograph), Jain's team can quantify the tunnelling rate of electrons out of the metal.

This experimental approach shows how to bypass the need to match the temperature distributions of the photocatalysis and thermocatalysis experiments to enable a correct identification of the underlying mechanism responsible for the catalysis~\cite{Y2-eppur-si-riscalda}. However, in other cases where the thermal control was performed under more severe conditions compared to that of the photocatalysis case (see e.g., Ref.~\citenum{Halas-Nature-Catalysis-2020}), the reaction rate was not negligible so that one cannot quantify the relative importance of non-thermal electrons to the reaction rate. In that sense, the conclusions of Ref.~\citenum{Halas-Nature-Catalysis-2020} are unfortunately invalid.

A possible reason for the significant non-thermal electron action in this study is the presence of hole scavengers, which effectively prevent the electron-hole pairs generated by the plasmonic decay from recombining. In bond-dissociation reactions (where no hole-scavengers were present, e.g. Refs.~\citenum{Halas_Science_2018,Halas_H2_dissociation_SiO2,plasmonic_photocatalysis_Linic}) the electron-hole pairs generated from plasmon decay recombine or thermalize at femto-second timescale~\cite{Dubi-Sivan}. However, when a hole-scavenger is present, the holes generated in the plasmon decay process are quickly filled by electrons from the hole-scavenger molecules in the solution. The high-energy non-thermal electrons, having their place taken by electrons from the hole scavengers, are trapped at high-energy states for much longer, sufficiently long so they can participate in the redox reaction. Thus, our thermal calculations re-affirm the conclusions in Ref.~\citenum{yu2019plasmonic}, and support a similar approach adopted earlier~\cite{JACS_Xiong,Yugang_Sun,yu2018plasmonic,Giulia_2019,Boltasseva_LPR_2020,Nguyen-Merced-redox}.

\section{Summary and outlook}
The thermal simulations performed in this manuscript demonstrate again that detailed thermal calculations can enable the correct identification of the underlying physical mechanism responsible for the speeding up of a certain chemical reaction by revealing the steady-state spatial distribution of the temperature  as before~\cite{anti-Halas-Science-paper,Y2-eppur-si-riscalda,R2R,Dubi-Sivan-APL-Perspective}, and here also by looking at the temperature dynamics. 

The next challenge in understanding the underlying physics/chemistry in such systems and quantifying it is to introduce aspects of the electron dynamics into the quantitative thermal model, and potentially even introduce a quantitative model of chemical interactions, electron transfer between the NPs and the chemical moieties and possibly reaction dynamics. This will require quantitative calculations of ``hot'' electrons (as in Ref.~\citenum{Dubi-Sivan,Dubi-Sivan-Faraday}) as well as of their tunnelling (as e.g., in Refs.~\citenum{Uriel_Schottky_2018,Giulia_Nat_Comm_2018,Khurgin-Faraday-hot-es}) and the chemical reaction dynamics itself. 

The formulation presented here  will also enable the quantitative study of electrochemical reactions, which attain growing attention in the context of plasmon-assisted photocatalysis, see, e.g. Refs.~\citenum{Tatsuma_2004,Willets_electrochemistry_2018,Rodriguez-2018,Caleb_Hill_Heating_Electrochemical_JPCC_2019,Cortes_Nano_lett_2019,Lange_2020}. We are hopeful that similar analysis can be applied to future experiments, in pursuit for the correct interpretation of the mechanism underlying the speeding up of the chemical reaction.

\begin{acknowledgement}
We wish to thank J. Huang, J. J. Baumberg and P. K. Jain for sharing their raw data and for many useful discussions; we also thank Y. Dubi for many useful conversations and Y. Green and Y. Feldman for their guidance on fluid mechanics simulations. We acknowledge funding from the Ben-Gurion University VP office for ``multi-disciplinary research''.
\end{acknowledgement}

%\begin{suppinfo}

%This will usually read something like: ``Experimental procedures and characterization data for all new compounds. The class will automatically add a sentence pointing to the information on-line:

%\end{suppinfo}

%\bibliography{my_bib}

\begin{mcitethebibliography}{66}
\bibitem[Khurgin(2019)]{Khurgin-Faraday-hot-es}
Khurgin,~J. Hot carriers generated by plasmons: where are they are generated
  and where do they go from there? \emph{Faraday Discussions} \textbf{2019},
  \emph{214}, 35--58\relax
\mciteBstWouldAddEndPuncttrue
\mciteSetBstMidEndSepPunct{\mcitedefaultmidpunct}
{\mcitedefaultendpunct}{\mcitedefaultseppunct}\relax
\EndOfBibitem
\end{mcitethebibliography}
\providecommand{\latin}[1]{#1}
\makeatletter
\providecommand{\doi}
  {\begingroup\let\do\@makeother\dospecials
  \catcode`\{=1 \catcode`\}=2 \doi@aux}
\providecommand{\doi@aux}[1]{\endgroup\texttt{#1}}
\makeatother
\providecommand*\mcitethebibliography{\thebibliography}
\csname @ifundefined\endcsname{endmcitethebibliography}
  {\let\endmcitethebibliography\endthebibliography}{}

\end{document}